\PassOptionsToPackage{unicode}{hyperref}
\PassOptionsToPackage{hyphens}{url}
\PassOptionsToPackage{dvipsnames,svgnames,x11names}{xcolor}
\documentclass[
]{article}

\usepackage{amsmath,amssymb}
\usepackage{iftex}
\ifPDFTeX
  \usepackage[T1]{fontenc}
  \usepackage[utf8]{inputenc}
  \usepackage{textcomp} 
\else 
  \usepackage{unicode-math}
  \defaultfontfeatures{Scale=MatchLowercase}
  \defaultfontfeatures[\rmfamily]{Ligatures=TeX,Scale=1}
\fi
\usepackage{lmodern}
\ifPDFTeX\else  
\fi
\IfFileExists{upquote.sty}{\usepackage{upquote}}{}
\IfFileExists{microtype.sty}{
  \usepackage[]{microtype}
  \UseMicrotypeSet[protrusion]{basicmath} 
}{}
\makeatletter
\@ifundefined{KOMAClassName}{
  \IfFileExists{parskip.sty}{%
    \usepackage{parskip}
  }{
    \setlength{\parindent}{0pt}
    \setlength{\parskip}{6pt plus 2pt minus 1pt}}
}{
  \KOMAoptions{parskip=half}}
\makeatother
\usepackage{xcolor}
\ifx\paragraph\undefined\else
  \let\oldparagraph\paragraph
  \renewcommand{\paragraph}[1]{\oldparagraph{#1}\mbox{}}
\fi
\ifx\subparagraph\undefined\else
  \let\oldsubparagraph\subparagraph
  \renewcommand{\subparagraph}[1]{\oldsubparagraph{#1}\mbox{}}
\fi

\usepackage{longtable,booktabs,array}
\usepackage{calc} 
\usepackage{etoolbox}
\makeatletter
\patchcmd\longtable{\par}{\if@noskipsec\mbox{}\fi\par}{}{}
\makeatother
\IfFileExists{footnotehyper.sty}{\usepackage{footnotehyper}}{\usepackage{footnote}}
\makesavenoteenv{longtable}
\usepackage{graphicx}
\makeatletter
\def\maxwidth{\ifdim\Gin@nat@width>\linewidth\linewidth\else\Gin@nat@width\fi}
\def\maxheight{\ifdim\Gin@nat@height>\textheight\textheight\else\Gin@nat@height\fi}
\makeatother
\setkeys{Gin}{width=\maxwidth,height=\maxheight,keepaspectratio}
\makeatletter
\def\fps@figure{htbp}
\makeatother
\newlength{\cslhangindent}
\newlength{\csllabelwidth}
\newlength{\cslentryspacingunit} 
\newenvironment{CSLReferences}[2] 
 {
  \setlength{\parindent}{0pt}
  \ifodd #1
  \let\oldpar\par
  \def\par{\hangindent=\cslhangindent\oldpar}
  \fi
  \setlength{\parskip}{#2\cslentryspacingunit}
 }%
 {}
\usepackage{calc}

\usepackage{booktabs}
\usepackage{longtable}
\usepackage{array}
\usepackage{multirow}
\usepackage{wrapfig}
\usepackage{float}
\usepackage{colortbl}
\usepackage{pdflscape}
\usepackage{tabu}
\usepackage{threeparttable}
\usepackage{threeparttablex}
\usepackage[normalem]{ulem}
\usepackage{makecell}
\usepackage{xcolor}
\usepackage{arxiv}
\usepackage{orcidlink}
\usepackage{amsmath}
\usepackage[T1]{fontenc}
\makeatletter
\@ifpackageloaded{caption}{}{\usepackage{caption}}
\AtBeginDocument{%
\ifdefined\contentsname
  \renewcommand*\contentsname{Table of contents}
\else
  \newcommand\contentsname{Table of contents}
\fi
\ifdefined\listfigurename
  \renewcommand*\listfigurename{List of Figures}
\else
  \newcommand\listfigurename{List of Figures}
\fi
\ifdefined\listtablename
  \renewcommand*\listtablename{List of Tables}
\else
  \newcommand\listtablename{List of Tables}
\fi
\ifdefined\figurename
  \renewcommand*\figurename{Figure}
\else
  \newcommand\figurename{Figure}
\fi
\ifdefined\tablename
  \renewcommand*\tablename{Table}
\else
  \newcommand\tablename{Table}
\fi
}
\@ifpackageloaded{float}{}{\usepackage{float}}
\floatstyle{ruled}
\@ifundefined{c@chapter}{\newfloat{codelisting}{h}{lop}}{\newfloat{codelisting}{h}{lop}[chapter]}
\floatname{codelisting}{Listing}

\makeatother
\makeatletter
\@ifpackageloaded{caption}{}{\usepackage{caption}}
\@ifpackageloaded{subcaption}{}{\usepackage{subcaption}}
\makeatother
\makeatletter
\@ifpackageloaded{tcolorbox}{}{\usepackage[skins,breakable]{tcolorbox}}
\makeatother
\makeatletter
\@ifundefined{shadecolor}{\definecolor{shadecolor}{rgb}{.97, .97, .97}}
\makeatother
\ifLuaTeX
  \usepackage{selnolig}  
\fi
\IfFileExists{bookmark.sty}{\usepackage{bookmark}}{\usepackage{hyperref}}
\IfFileExists{xurl.sty}{\usepackage{xurl}}{} 
\hypersetup{
  pdftitle={biorecap: an R package for summarizing bioRxiv preprints with a local LLM},
  pdfauthor={Stephen D. Turner},
  pdfkeywords={Large language models, LLM, Science
communication, Preprints, R package},
  colorlinks=true,
  linkcolor={blue},
  filecolor={Maroon},
  citecolor={Blue},
  urlcolor={Blue},
  pdfcreator={LaTeX via pandoc}}

\renewcommand{\today}{August 21, 2024}
\newcommand{\runninghead}{A Preprint }
\renewcommand{\runninghead}{biorecap }
\title{biorecap: an R package for summarizing bioRxiv preprints with a
local LLM}
\author{\textbf{Stephen D.
Turner}~\orcidlink{0000-0001-9140-9028}\\\\Citizen
Scientist\\Charlottesville,
Virginia\\\href{mailto:https://stephenturner.us}{https://stephenturner.us}}
\date{}
\begin{document}
\maketitle
\begin{abstract}
The establishment of bioRxiv facilitated the rapid adoption of preprints
in the life sciences, accelerating the dissemination of new research
findings. However, the sheer volume of preprints published daily can be
overwhelming, making it challenging for researchers to stay updated on
the latest developments. Here, I introduce \texttt{biorecap}, an R
package that retrieves and summarizes bioRxiv preprints using a large
language model (LLM) running locally on nearly any commodity laptop.
\texttt{biorecap} leverages the \texttt{ollamar} package to interface
with the Ollama server and API endpoints, allowing users to prompt any
local LLM available through Ollama. The package follows tidyverse
conventions, enabling users to pipe the output of one function as input
to another. Additionally, \texttt{biorecap} provides a single wrapper
function that generates a timestamped CSV file and HTML report
containing short summaries of recent preprints published in
user-configurable subject areas. By combining the strengths of LLMs with
the flexibility and security of local execution, \texttt{biorecap}
represents an advancement in the tools available for managing the
information overload in modern scientific research. The
\texttt{biorecap} R package is available on GitHub at
\url{https://github.com/stephenturner/biorecap} under an open-source
(MIT) license.
\end{abstract}
{\bfseries \emph Keywords}
\def\sep{\textbullet\ }
Large language models \sep LLM \sep Science
communication \sep Preprints \sep 
R package

\ifdefined\Shaded\renewenvironment{Shaded}{\begin{tcolorbox}[borderline west={3pt}{0pt}{shadecolor}, enhanced, breakable, boxrule=0pt, frame hidden, sharp corners, interior hidden]}{\end{tcolorbox}}\fi

\hypertarget{sec-intro}{%
\section{Introduction}\label{sec-intro}}

Large language models (LLMs) have revolutionized the way we process and
interact with large amounts of text data. These models show remarkable
capabilities in generating human-like text, summarizing documents,
translating languages, and performing complex analyses across various
domains. Until recently, the best-in-class ``frontier'' models only
included closed-source, closed-weight models: GPT-4 and GPT4o from
OpenAI, Gemini from Google, and Claude 3.5 Sonnet from Anthropic. Aside
from perpetual subscription costs required to use these models, reliance
on cloud-based systems could raise concerns about data privacy,
security, and the potential misuse of sensitive information,
particularly in fields such as healthcare and biotechnology.

The emergence of open models such as the Llama 3 herd of models (Dubey
et al. 2024), marks a significant shift in how these powerful tools can
be used. Llama 3.1 405B (405 billion parameters) is competitive with the
top frontier models listed above in knowledge, math, tool use, coding,
multilingual translation, and enhanced reasoning. The smaller models
including an 8B and 70B versions are small and fast enough to run on a
commodity laptop, and routinely outperform the previous generation of
frontier models (GPT 3.5, Gemini 1.0 Pro, Claude 3 Sonnet) in human
evaluations (Zheng et al. 2023).

Locally hosted models offer several advantages, including enhanced
control over data, reduced latency in processing, and the ability to
operate in environments with restricted or no internet access.
Additionally, being able to run lightweight models on a commodity laptop
or inexpensive cloud VM reduces or eliminates ongoing costs. These
benefits make locally running LLMs particularly appealing for academic
research, where funds are limited and the integrity and confidentiality
of data are paramount. Moreover, the accessibility of open models
democratizes the use of LLMs, enabling researchers and developers to
customize and optimize these tools for specific applications without the
constraints imposed by proprietary platforms.

One of the most pressing challenges in modern scientific research is the
ability to keep pace with the rapid dissemination of new findings. The
establishment and rise of the bioRxiv preprint server (Sever et al.
2019) for life sciences has accelerated the rate at which new research
becomes available, often outpacing the capacity of researchers to stay
updated. While ultimately beneficial for the advancement of science,
this flood of new research published daily can be overwhelming, leading
to difficulties in identifying the most relevant studies and
synthesizing the latest developments in a timely manner. This paper
introduces \texttt{biorecap}, an R package designed to help researchers
summarize the large number of preprints published daily in the life
sciences.

\hypertarget{implementation}{%
\section{Implementation}\label{implementation}}

\texttt{biorecap} is an R package that retrieves and summarizes bioRxiv
preprints using a local LLM using the \texttt{ollamar} package (Lin
2024) as an interface to Ollama (\url{https://ollama.com/}) server and
API endpoints. The \texttt{biorecap} package first retrieves recent
bioRxiv preprints in user-specified subject areas using bioRxiv's RSS
feed. \texttt{biorecap} will then construct a prompt to summarize each
paper based on the title and abstract in a user-specified number of
sentences, then queries a selected local LLM to produce a summary of
each paper.

\texttt{biorecap} can prompt any local LLM available through Ollama,
such as Llama 3.1 from Meta (Dubey et al. 2024), Gemma2 from Google
(Gemma Team et al. 2024), Qwen2 from Alibaba (Bai et al. 2023), Mistral
from Mistral AI (Jiang et al. 2023), Phi-3 from Microsoft (Abdin et al.
2024), and many others.

The \texttt{biorecap} package follows tidyverse conventions, where the
output of one function can be piped as input to another:
\texttt{get\_preprints()\ \textbar{}\textgreater{}\ add\_prompts()\ \textbar{}\textgreater{}\ add\_summary()}.
Finally, the \texttt{biorecap\_report()} function takes as input one or
more subject areas, the name of a local LLM to use, creates summaries of
all recent preprints in these subject areas, and creates an HTML report
based on a parameterized RMarkdown template built into the package
itself.

\hypertarget{results}{%
\section{Results}\label{results}}

Running \texttt{get\_preprints()} will fetch the latest preprints from
bioRxiv for any supplied subject, returning titles, abstracts and URLs.
Table~\ref{tbl-pp} shows the results from running
\texttt{get\_preprints()} for bioinformatics, genomics, and synthetic
biology.

\hypertarget{tbl-pp}{}
\begin{table}
\caption{\label{tbl-pp}Results from running \texttt{get\_preprints()} on select subjects. This
function pulls the latest preprints from bioRxiv for a user-supplied
vector of subjects. URL column omitted, abstract column truncated, and
only a subset of rows is shown for each subject. These results were
collected from the bioRxiv RSS feeds on August 6, 2024. }\tabularnewline

\centering
\begin{tabular}[t]{l>{\raggedright\arraybackslash}p{20em}>{\raggedright\arraybackslash}p{15em}}
\toprule
subject & title & abstract\\
\midrule
\cellcolor{gray!10}{bioinformatics} & \cellcolor{gray!10}{SeuratExtend: Streamlining Single-Cell RNA-Seq Analysis Through an Integrated and Intuitive Framework} & \cellcolor{gray!10}{Single-cell RNA sequencing (scRNA-seq) has revolutionized the study of cellular heterog...}\\
bioinformatics & An Evolutionary Statistics Toolkit for Simplified Sequence Analysis on Web with Client-Side Processing & We present the "Evolutionary Statistics Toolkit", a user-friendly web-based platform de...\\
\cellcolor{gray!10}{bioinformatics} & \cellcolor{gray!10}{A map of integrated cis-regulatory elements enhances gene regulatory analysis in maize} & \cellcolor{gray!10}{Cis-regulatory elements (CREs) are non-coding DNA sequences that modulate gene expressi...}\\
bioinformatics & ... & ...\\
\cellcolor{gray!10}{genomics} & \cellcolor{gray!10}{Nanopore-based analysis unravels the genetic landscape and phylogenetic placement of human-infecting Trichuris species in Cote d'Ivoire, Tanzania, Uganda, and Laos} & \cellcolor{gray!10}{Soil-transmitted helminthiases (STH), including trichuriasis, pose a significant global...}\\
\addlinespace
genomics & EPInformer: A Scalable Deep Learning Framework for Gene Expression Prediction by Integrating Promoter-enhancer Sequences with Multimodal Epigenomic Data & Transcriptional regulation, critical for cellular differentiation and adaptation to env...\\
\cellcolor{gray!10}{genomics} & \cellcolor{gray!10}{Complete sequencing of ape genomes} & \cellcolor{gray!10}{We present haplotype-resolved reference genomes and comparative analyses of six ape spe...}\\
genomics & ... & ...\\
\cellcolor{gray!10}{synthetic\_biology} & \cellcolor{gray!10}{Colorimetric CRISPR Biosensor: A Case Study with Salmonella Typhi} & \cellcolor{gray!10}{There is a critical need to implement a sensitive and specific point-of-care (POC) bios...}\\
synthetic\_biology & Accelerated enzyme engineering by machine-learning guided cell-free expression & Enzyme engineering is limited by the challenge of rapidly generating and using large da...\\
\addlinespace
\cellcolor{gray!10}{synthetic\_biology} & \cellcolor{gray!10}{Establishment of a rapid method to assemble and transfer DNA fragments into the JCVI-syn3B minimal synthetic bacterial genome} & \cellcolor{gray!10}{JCVI-syn3B (syn3B), a minimal synthetic bacterium that only possesses essential genes, ...}\\
synthetic\_biology & ... & ...\\
\bottomrule
\end{tabular}
\end{table}

The \texttt{add\_prompt()} adds a prompt column. Each row will contain a
preamble (instructions to the model, such as the number of sentences to
summarize the paper in and other user-configurable instructions), along
with the title and abstract. An example prompt for the first paper shown
in Table~\ref{tbl-pp} is shown below (abstract truncated).

\begin{quote}
\textbf{Prompt}:

I am giving you a paper's title and abstract. Summarize the paper in as
many sentences as I instruct. Do not include any preamble text to the
summary just give me the summary with no preface or intro sentence.

Number of sentences in summary: 2

Title: SeuratExtend: Streamlining Single-Cell RNA-Seq Analysis Through
an Integrated and Intuitive Framework

Abstract: Single-cell RNA sequencing (scRNA-seq) has revolutionized the
study of cellular heterogeneity, but the rapid expansion of analytical
tools has proven to be both a blessing and a curse, presenting
researchers with significant challenges. Here, we present SeuratExtend
\ldots(truncated)\ldots{}
\end{quote}

This prompt is for the SeuratExtend preprint (Hua et al. 2024). The
\texttt{add\_summary()} function will generate a response from the
specified model for the prompt in each row, adding a summary column to
the data frame. Results from prompting Llama3.1:8b with this prompt is
shown below.

\begin{quote}
\textbf{Summary}: SeuratExtend is an R package that integrates essential
tools and databases for single-cell RNA sequencing (scRNA-seq) data
analysis, streamlining the process through a user-friendly interface.
The package offers various analyses, including functional enrichment and
gene regulatory network reconstruction, and seamlessly integrates
multiple databases and popular Python tools.
\end{quote}

\begin{figure}

{\centering \includegraphics{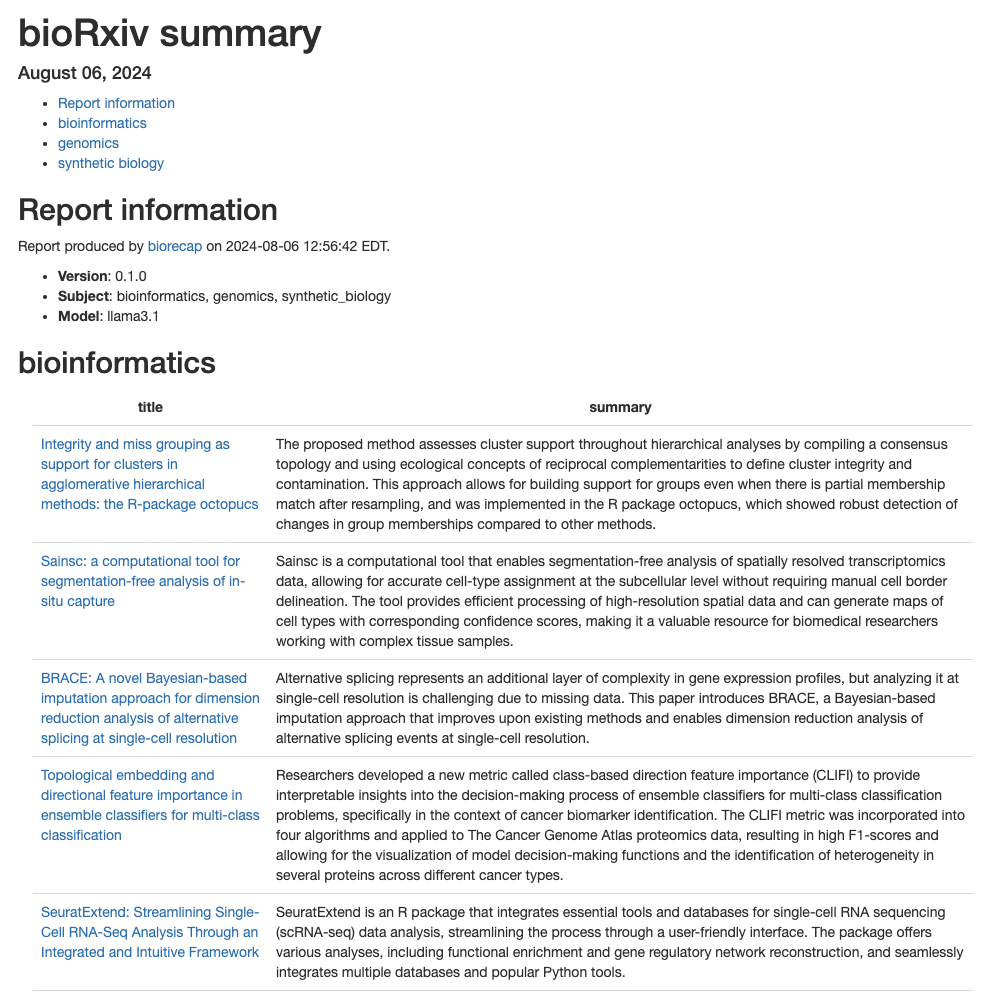}

}

\caption{\label{fig-report}Example biorecap report for bioinformatics,
genomics, and synthetic biology from August 6, 2024.}

\end{figure}

Finally, the \texttt{biorecap} package provides another function,
\texttt{biorecap\_report()} which runs all of these functions
successively, writing the resulting data frame to a CSV file, and
producing an HTML report from a parameterized RMarkdown template
installed with the package. An excerpt from an example report is shown
in Figure~\ref{fig-report}.

\hypertarget{discussion}{%
\section{Discussion}\label{discussion}}

The \texttt{biorecap} package provides an API to retrieve recent
preprints from bioRxiv and summarize them using a local LLM. By
combining the strengths of LLMs with the flexibility and security of
local execution, \texttt{biorecap} represents an advancement in the
tools available for managing the information overload in modern
scientific research. Using the built-in parameterized RMarkdown report
template, a single function, \texttt{biorecap\_report()} can produce an
HTML report containing short summaries of recent papers published in a
user-defined collection of subjects. It should be possible to automate
the creation of these reports with a lightweight orchestrator such as
the maestro R package (Hipson and Garnett 2024), or through a simple
cron job.

The \texttt{biorecap} package has some notable limitations. First, the
number of preprints returned with \texttt{get\_preprints()} is limited
by the number of preprints available in bioRxiv's RSS feeds. Currently,
this is limited to 30 preprints per subject area. Additionally, while
the initial version of \texttt{biorecap} was designed with bioRxiv in
mind, a future version will include the ability to query medRxiv
preprints as well (\url{https://www.medrxiv.org/}). Finally, a future
version will incorporate the ability to summarize all recent papers in a
subject to produce a top line summary on interesting developments in a
particular area each day.

The \texttt{biorecap} R package is available on GitHub at
\url{https://github.com/stephenturner/biorecap} under an open-source
(MIT) license.

\newpage{}

\newpage{}

\newpage{}

\hypertarget{references}{%
\section*{References}\label{references}}
\addcontentsline{toc}{section}{References}

\hypertarget{refs}{}
\begin{CSLReferences}{1}{0}
\leavevmode\vadjust pre{\hypertarget{ref-abdin2024}{}}%
Abdin, Marah, Sam Ade Jacobs, Ammar Ahmad Awan, Jyoti Aneja, Ahmed
Awadallah, Hany Awadalla, Nguyen Bach, et al. 2024. {``Phi-3 Technical
Report: A Highly Capable Language Model Locally on Your Phone.''}
\url{https://doi.org/10.48550/ARXIV.2404.14219}.

\leavevmode\vadjust pre{\hypertarget{ref-bai2023}{}}%
Bai, Jinze, Shuai Bai, Yunfei Chu, Zeyu Cui, Kai Dang, Xiaodong Deng,
Yang Fan, et al. 2023. {``Qwen Technical Report.''}
\url{https://doi.org/10.48550/ARXIV.2309.16609}.

\leavevmode\vadjust pre{\hypertarget{ref-dubey2024a}{}}%
Dubey, Abhimanyu, Abhinav Jauhri, Abhinav Pandey, Abhishek Kadian, Ahmad
Al-Dahle, Aiesha Letman, Akhil Mathur, et al. 2024. {``The Llama 3 Herd
of Models.''} \url{https://doi.org/10.48550/ARXIV.2407.21783}.

\leavevmode\vadjust pre{\hypertarget{ref-gemmateam2024}{}}%
Gemma Team, Thomas Mesnard, Cassidy Hardin, Robert Dadashi, Surya
Bhupatiraju, Shreya Pathak, Laurent Sifre, et al. 2024. {``Gemma: Open
Models Based on Gemini Research and Technology.''}
\url{https://doi.org/10.48550/ARXIV.2403.08295}.

\leavevmode\vadjust pre{\hypertarget{ref-maestro}{}}%
Hipson, Will, and Ryan Garnett. 2024. {``Maestro: Orchestration of Data
Pipelines.''} \url{https://CRAN.R-project.org/package=maestro}.

\leavevmode\vadjust pre{\hypertarget{ref-hua2024}{}}%
Hua, Yichao, Linqian Weng, Fang Zhao, and Florian Rambow. 2024.
{``SeuratExtend: Streamlining Single-Cell RNA-Seq Analysis Through an
Integrated and Intuitive Framework.''}
\url{http://dx.doi.org/10.1101/2024.08.01.606144}.

\leavevmode\vadjust pre{\hypertarget{ref-jiang2023}{}}%
Jiang, Albert Q., Alexandre Sablayrolles, Arthur Mensch, Chris Bamford,
Devendra Singh Chaplot, Diego de las Casas, Florian Bressand, et al.
2023. {``Mistral 7B.''} \url{https://doi.org/10.48550/ARXIV.2310.06825}.

\leavevmode\vadjust pre{\hypertarget{ref-ollamar}{}}%
Lin, Hause. 2024. {``Ollamar: 'Ollama' Language Models.''}
\url{https://CRAN.R-project.org/package=ollamar}.

\leavevmode\vadjust pre{\hypertarget{ref-sever2019}{}}%
Sever, Richard, Ted Roeder, Samantha Hindle, Linda Sussman, Kevin-John
Black, Janet Argentine, Wayne Manos, and John R. Inglis. 2019.
{``bioRxiv: The Preprint Server for Biology.''}
\url{http://dx.doi.org/10.1101/833400}.

\leavevmode\vadjust pre{\hypertarget{ref-zheng2023}{}}%
Zheng, Lianmin, Wei-Lin Chiang, Ying Sheng, Tianle Li, Siyuan Zhuang,
Zhanghao Wu, Yonghao Zhuang, et al. 2023. {``LMSYS-Chat-1M: A
Large-Scale Real-World LLM Conversation Dataset.''}
\url{https://doi.org/10.48550/ARXIV.2309.11998}.

\end{CSLReferences}

\end{document}